\documentclass[12pt,preprint]{aastex}

\newcommand{\be}{\begin{equation}}
\newcommand{\ee}{\end{equation}}

\def\lta{\,\raise 0.3 ex\hbox{$ < $}\kern -0.75 em
 \lower 0.7 ex\hbox{$\sim$}\,}
\def\gta{\,\raise 0.3 ex\hbox{$ > $}\kern -0.75 em
 \lower 0.7 ex\hbox{$\sim$}\,}

\newcommand{\simless}{\lower.5ex\hbox{$\; \buildrel < \over \sim\;$}}
\newcommand{\simgreat}{\lower.5ex\hbox{$\; \buildrel > \over \sim\;$}} 

\begin{document}

\title{Star Formation at the Galactic Center}

\author{Marco Fatuzzo,$^1$ and Fulvio Melia$^{2}$} 
 
\affil{$^1$Physics Department, Xavier University, Cincinnati, OH 45207} 

\affil{$^2$Department of Physics and Steward Observatory, The University 
of Arizona, Tucson, Arizona 85721}

\email{fatuzzo@xavier.edu; melia@physics.arizona.edu}

\begin{abstract}
Molecular clouds at the Galactic center (GC) have environments 
considerably different from their disk counterparts. 
The GC may therefore provide important clues about how the
environment affects star formation. Interestingly, while the 
inner 50 parsecs of our Galaxy include a remarkable population of 
high-mass stars, the initial mass function (IMF) 
appears to be consistent with a Salpeter slope down to 
$\sim 1\;M_\odot$.   We show here that the loss of turbulent 
pressure due to ambipolar diffusion and the damping of Alfv\'en and fast MHD 
waves can lead to the formation of dense 
condensations exceeding their Jeans limit. The fragmentation and subsequent
collapse of these condensations is similar to the diffusion-driven 
protostellar collapse mechanism expected to occur within nearby ``regular" 
molecular clouds.  As such, a Salpeter IMF at the GC is not surprising,
though the short dynamical timescales associated with the GC molecular clouds
may help explain the lower star formation efficiency observed from this region.
\end{abstract}

\begin{keywords}
{ISM: Clouds, Magnetic Fields, Stars: formation, Turbulence, Galaxy: Centre, Stellar Content}
\end{keywords}

\section{Introduction}
Basic star formation theory holds that a star's initial mass is dependent, 
at least in part, on the environment in which it is born (Bonnell et al. 2007).
It is quite reasonable therefore to expect that the IMF may vary for 
different types of stellar populations. Yet the IMF appears to be quite 
uniform throughout our galaxy (Kroupa 2002; Massey 2003), being well described
by a power-law form $dN / dm \propto m^\alpha$ for $m > 1\;M_\odot$ and a
lognormal form below this value (Adams \& Fatuzzo 1996; Kroupa 2001; Chabrier 2003).
At present, there is only one known cluster in our Galaxy that appears to have a
considerably top-heavy IMF---the GC cluster. 
Two other clusters that lie within 50 pc in projection from
the nucleus also contain a remarkable number of high-mass stars---the
Quintuplet and the Arches clusters. The Quintuplet cluster
contains more Wolf-Rayet stars than any other cluster in the galaxy. However,
it is significantly more dispersed than the Arches
cluster, and as such, its mass function has not been determined.
In contrast to the Quintuplet, the Arches cluster is quite dense,
containing at least 150 O stars within a radius of
0.6 pc.  While the mass
function for this cluster was initially thought to be 
considerably shallower than a Salpeter MF (Stolte et al. 2005; Kim et al. 2006),
the peculiarities in the Arches mass function can
apparently be attributed to evolution, and the Arches initial mass function may well
be consistent with a Salpeter slope down to $\sim 1\;M_\odot$ (Portegies Zwart et al. 2007).  

These results are somewhat surprising given that the 
molecular clouds within $\sim 100$ pc of the GC have considerably different 
environments  than their disk counterparts (e.g., Crocker et al. 2007).
While a lot of progress has been made over the past decade in the field
of star formation theory, a complete understanding of how a star acquires its initial mass 
remains elusive. At present, two different star formation paradigms are seemingly 
favored by the star formation community---the ``standard" paradigm (Shu et al. 1987; see also
more recent works by Adams \& Shu 2007; Tassis \& Mouschovias 2007; Kudoh \&
Basu 2008; Basu et al. 2009)
and the turbulent fragmentation paradigm (Padoan \& Nordlund 2002; 
Mac Low \& Klessen 2004; Bonnell et al. 2007).
On the observational front, a recent survey
of magnetic field strengths in nearby dark cloud cores aimed at testing
these scenarios failed to yield conclusive results (Troland \& Crutcher 2008). 
The extreme molecular cloud environments at the GC (Yusef-Zadeh et al. 2000; Melia
\& Falcke 2001)
may therefore provide an important test case from which to consider star formation
theory.  Indeed, recent observations now make it possible to consider certain aspects
of star formation within the context of the standard paradigm. 

For example, Yusef-Zadeh et al. (2007) have recently compiled evidence for an 
enhancement of cosmic-ray electrons within the Galactic center region, and claim 
that this enhancement may be responsible for heating the gas clouds to temperatures 
above the dust temperature.  These authors note that higher cloud temperatures 
increase the Jeans mass, thereby accounting for what at the time was believed to 
be a top-heavy IMF at the Galactic Center (Stolte 2005; Kim et al. 2006).  In addition, 
the authors suggest that the high ionization, which would increase magnetic 
coupling to the cloud material and lengthen the ambipolar diffusion time, could 
be responsible for the low star formation efficiency observed in this region
(Gordon et al. 1993; Lis et al. 2001). 

We extend this line of investigation by considering how the extreme environments 
associated with the dense cores of the GC molecular clouds impact star formation 
within those regions.  Given the strengths of the magnetic fields 
which permeate this region, the critical mass below which cores can be supported
against gravitational collapse is quite a bit higher in this region than
in typical disk GMCs. While the high ionization rates at the Galactic Center
increase the coupling between the magnetic field and neutral gas, as noted
by Yusef-Zadeh et al. (2007), the high densities of molecular
gas in this region offsets the higher ionization
rate's effect on the ambipolar diffusion time, which scales
as $\tau_{AD}\propto (\zeta/n)^{1/2}$.  As a result, magnetic support
can be effectively removed in the densest regions of the GC molecular clouds,
consistent with the observation that the most active star formation regions in the nuclear 
disk, such as Sgr B2, are associated with a dense environment.  

Interestingly, magnetic support within the dense regions of the GC molecular clouds
can also be efficiently removed via strong damping of Alfv\'en and fast MHD waves,
a process that has been proposed for removing part of the support
against gravity and thereby initiating fragmentation
(or core formation) in typical molecular clouds (Mouschovias
1987; Mouschovias \& Psaltis 1995).
A second important aspect of this work is the assessment of what impact 
this mechanism has on the loss of magnetic support in the GC molecular
clouds.  Specifically, while 
ambipolar diffusion acts to gradually remove magnetic support in massive dense cores,
the reduction of magnetic pressure within 
the densest gas of the GC may result preferentially from 
MHD wave damping (Mouschovias 1991, Balsara 1996), which would then lead to 
the formation of dense condensations
with masses that, depending on environmental conditions, 
can range between $\sim 0.1$--$30\;M_\odot$.
Although condensations whose masses exceed their Jeans limit may initially
be supported against gravitational collapse 
by slow MHD waves, the relatively short ambipolar
diffusion times associated with these objects (given their greater density
over the surrounding medium) would efficiently remove
the remaining magnetic pressure support.

\section{Impact of the GC Environment on Star Formation}

The molecular clouds near the GC have environments considerably
different from their galactic disk counterparts.  
For example, the average molecular hydrogen number density $\langle n_{H_2} \rangle$
over the Sgr B complex---the largest molecular cloud complex (Lis \& Goldsmith 1989; 
Lis \& Goldsmith 1990; Paglione et al. 1998)
near the GC---has a density $3$--$10 \times 10^3$ cm$^{-3}$.
Like its traditional counterparts, Sgr B displays a highly nonlinear structure,
containing two bright sub-regions, Sgr B1 and Sgr B2, the latter having
an average molecular density of $\sim 10^6$ cm$^{-3}$, and containing
three dense ($n \sim 10^{7.3-8}$ cm$^{-3}$), small ($r\sim 0.1$ pc) 
cores---labeled North, Main and South.
These cores also show considerable structure,
containing numerous ultra-compact and hyper-compact
HII regions.  As such, the densities associated with the GC molecular clouds
are about two orders of magnitude greater than those in the disk of our galaxy.

In the simplest strong-magnetic field case
(which is the focus of our study), flux freezing arguments imply
the magnetic field strength $B$ in the interstellar
medium scales to the gas density as $B\propto n^{1/2}$.
Indeed, an analysis of magnetic field strengths measured
in molecular clouds by Crutcher (1999) yielded a relation between $B$ and $n$
of the form $B \propto n^{0.47}$, although there is a significant 
amount of scatter in the data
used to produce this fit.  
Interestingly, one of the data points used in establishing 
this relation comes from observations of Sgr B2 (Crutcher 1996), 
for which $B_{los} = 0.5$ mG at a density of $n = 2,500$ cm$^{-3}$.  
This datum is an outlier to the rest of the data, being 
well-above the established fitted scaling relation.  
A significantly better correlation to the data was obtained in a 
subsequent analysis by Basu (2000) who, by including the effects of cloud 
flattening along the mean magnetic field direction, obtained the relation 
\begin{equation}
B = (8\pi c_1)^{1/2} \,{\sigma_v\,\rho^{1/2}\over\mu}\,,
\end{equation}
where $c_1 \simgreat 1$ is an undetermined proportionality constant between the mean 
mass density $\rho$ and the midplane mass density, $\sigma_v$ is the total one-dimensional
velocity dispersion (related to the FWHM of spectral lines $\Delta V$ via the relation
$\sigma_v = \Delta V / [8\, {\rm ln} 2]^{1/2}$), and $\mu$ is the dimensionless
mass-to-flux ratio in units of the critical value $(2\pi G^{1/2})^{-1}$ for 
a disk. The best least-squares fit was obtained for $\sqrt{c_1}/\mu = 0.8$,
for which the Sgr B2 datum is no longer an outlier. This result agrees with the 
prediction of ambipolar diffusion driven star formation, and is consistent with the
idea that nonthermal linewidths arise from MHD fluctuations 
for which the Alfv\'en Mach number is of order unity. Specifically, MHD waves would
propagate through this medium with an Alfv\'en speed
\begin{equation}
v_\alpha = {B\over (4 \pi \rho)^{1/2}} = 31 \; {\rm km}\;{\rm s}^{-1}
\left({B\over 1 \,{\rm mG}}\right)\,
\left({n \over 2500\, {\rm cm}^{-3}}\right)^{-1/2}\,,
\end{equation} 
consistent with the interpretation that the observed 15--50 km s$^{-1}$ 
supersonic internal velocity dispersions (Morris \& Serabyn 1996) result 
from magnetic turbulence.

The coupling between the magnetic field and the molecular gas depends 
sensitively on the ion fraction, and hence, on the local ionization rate.  
Several recent studies indicate an excess of cosmic ray flux within 
the central region of the galaxy.  For example, the ionization rate 
for Sgr B has been determined to be $\sim 1$--$4\times 10^{-16}$ 
s$^{-1}$ (van der Takk et al. 1996), about an order of magnitude 
greater than the value determined for local molecular 
clouds (van der Tak \& van Dishoeck 2000).

Although the temperature structure in the GC molecular clouds is quite 
complicated, at least two kinetic temperatures are indicated, $T_{kin}\sim 200$ K and
$T_{kin}\sim 25$ K in each cloud, with evidence that the cooler gas
is associated with higher densities (Hu\"ttemeister et al. 1993b). Sgr B2 Main---with a 
temperature of $\sim 200$ K and density $\sim 2\times 10^7$ cm$^{-3}$---appears 
to be an exception (Hu\"ttemeister 1993a; De Vicente er al. 1996).  We note, however,  that
this structure contains numerous ultra-compact HII regions (de Pree et al. 1998),
so the high temperatures may be due to the presence of massive stars.

In the ``standard" star formation paradigm, dense cores are initially subcritical
and thus supported by magnetic turbulence until ambipolar diffusion
increases the mass to magnetic flux ratio sufficiently that collapse can
ensue.  This phase is then followed
by an inside out collapse that leads to the formation of a central
star. The characteristic ambipolar diffusion timescale under
quasi-static, subcritical assumptions is $\tau_{AD} \approx \tau_{ff}^2/\tau_{ni}$,
where $\tau_{ff} = \sqrt{3\pi/(32 G \rho)}$ and 
$\tau_{ni} \approx (\langle \sigma v \rangle_{in} n_i)^{-1}$ are, respectively, 
the free-fall time and the neutral-ion collision time (Mouschovias 1976; Shu 1983;
Shu 1992),
and $\langle \sigma v \rangle_{in}\approx 1.7\times 10^{-9}$ cm$^{-3}$ s$^{-1}$
is the average collision rate between ions and neutrals (Mouschovias 1991).
The ion density is well approximated by the relation
\begin{equation}
n_i = 3.2\times 10^{-3} \,{\rm cm}^{-3}\,\left({n\over 10^5 
\, {\rm cm}^{-3}}\right)^{1/2} \left({\zeta\over\zeta_{CR}}\right)^{1/2}\,,
\end{equation}
where $\zeta_{CR} = 10^{-17}$ s$^{-1}$, so long as $\zeta / n > 10^{-22} - 10^{-24}$
cm$^3$ s$^{-1}$ (Elmegreen 1979). Given that
$\zeta / n \sim 10^{-23}$ cm$^3$ s$^{-1}$ in the dense regions of the GC, we 
consider both the case that $n_i$ is given by Eq. (3) for all
values of $n$ (Case I), and the case where this relation holds only for
$n < n_{crit} = 10^6 \,(\zeta/\zeta_{CR})$ cm$^{-3}$, beyond which 
$n_i = 10^{-2}\, (\zeta/\zeta_{CR})$ cm$^{-3}$ (Case II).

Correspondingly, the ambipolar diffusion time is given by the expressions

$$
\tau_{AD} \approx 2.2 \,{\rm Myr}\, 
\left({\zeta\over \zeta_C}\right)^{1/2}\,
\left({n\over 10^5 \,{\rm cm}^{-3}}\right)^{-1/2}\,\eqno(4{\rm a})
$$
for Case I and for Case II when $n < n_{crit}$; and
$$
\tau_{AD}\approx 7.1 \,{\rm Myr}\, 
\left({\zeta\over \zeta_C}\right)\,
\left({n\over 10^5 \,{\rm cm}^{-3}}\right)^{-1}\,
\eqno(4{\rm b})
$$
for Case II when $n > n_{crit}$, in good agreement with results from more 
sophisticated analyses (Ciolek \& Mouschovias 1995). As pointed out by numerous authors,
this timescale is too slow, by a factor of $\sim$ 3--10, to account
for the observed statistics of starless molecular cores in typical 
molecular cloud environments (Jijina et al. 1999; but see also Mouschovias et al. 2006).  
However, several mechanisms 
have recently been proposed which speed up the process (Ciolek \& Basu 2001;
Zweibel 2002; Fatuzzo \& Adams 2002), 
thereby addressing one of the main criticism of the ``standard" paradigm.  

Dense cores ($r \sim 0.1$ pc, $n\sim 5\times 10^7$ cm$^{-3}$) within 
the GC molecular clouds contain several 
thousand solar masses of gas, well in excess of their Jeans limit.  
It would thus appear that magnetic turbulence is responsible for the
global support of these regions.  However, Alfv\'en and fast MHD waves can only couple
to the neutral gas if the ion-neutral collision time
is shorter than the MHD time. As a result, star formation can 
be initiated by the loss of turbulent support attributed to ambipolar diffusion
damping of hydromagnetic waves (Mouschovias 1987).  
This loss of turbulent support occurs below a lengthscale

$$
\lambda_\alpha \equiv \pi v_\alpha\tau_{ni}
\approx 1.8\times 10^{17} {\rm cm}\,
\left({v_\alpha\over 31 \,{\rm km/s}}\right)\,
\left({n\over 10^7\,{\rm cm}^{-3}}\right)^{-1/2}\,
\left({\zeta\over\zeta_{CR}}\right)^{-1/2}\,
\eqno(5{\rm a})
$$
for Case I and for Case II when $n < n_{crit}$; and
$$
\lambda_\alpha \approx 5.7\times 10^{17} {\rm cm}\,
\left({v_\alpha\over 31 \,{\rm km/s}}\right)\,
\left({\zeta\over\zeta_{CR}}\right)^{-1}\,
\eqno(5{\rm b})
$$
for Case II when $n > n_{crit}$, such that magnetic disturbances
with wavelengths $\lambda < \lambda_\alpha$
diffuse before collisions between the neutrals and
ions have had time to transmit to the neutrals the magnetic
force associated with the disturbance (Mouschovias 1991; Balsara 1996).
Below this lengthscale, Alfv\'en and fast MHD waves become
completely non-propagating and are quickly
damped.   

For molecular clumps in typical GMCs ($n \approx 10^{3}$ cm$^{-3}$; $v_\alpha 
\approx 1 $ km s$^{-1}$; $\zeta = \zeta_{CR}$),
$\lambda_\alpha \approx 0.2$ pc, smaller than the typical clump
size ($\sim 0.2 - 2$ pc) required for MHD waves to support these structures from collapse.
The loss of magnetic support on scales less than $\lambda_\alpha$ is therefore
expected to play a role in setting the length scale at which ambiplar diffusion
becomes significant and fragmentation or core formation is initiated 
(Mouschovias 1987; 1991).  In turn, there is a natural mass scale of $\sim 1 M_\odot$
associated with the ambipolar-driven star formation process.  
Indeed, massive clumps observed in typical GMCs are comprised of
$\sim 100$--$1000$ small ($R \sim 0.1 - 0.2$ pc), dense ($\sim 10^4$--$10^5$ cm$^{-3}$)
cores whose mass function has been measured to range from $\sim 1 - 100 M_\odot$
(with a peak $\sim 10 M_\odot$) by Jijina et al. (1999), and more recently,
from $\sim 0.2 - 20 M_\odot$ (with a characteristic mass of 
$\approx 2 M_\odot$) by Lada et al. (2008). 
We note, however, that while the loss of magnetic support due to wave damping
may serve to seed this structure, the evolutionary timescale for the 
collapse is still set by the ambipolar diffusion time rather than the free-fall time
(Mouschovias \& Psaltis 1995).
Furthermore,  structures in molecular clouds (e.g., cores) typically exhibit Bonnor-Ebert
density profiles.  That it, their densities scale as $r^{-2}$ until
flattening occurs at radii below $r_C \sim 10^{16}$ cm.
As a result, $\lambda_\alpha \propto n^{-1/2} \propto r$, and remains
smaller than the core size at a given density for all but the central
(flattened) part of the cores.\footnote{ 
Oishi and Mac Low (2006) find that
ambipolar diffusion is unable to set a characteristic scale for gravitational 
collapse and star formation in turbulent molecular clouds, owing presumably
to support from slow MHD waves.  However,
the simulations of Oishi and Mac Low (2006) were
performed in the absence of gravity, the presence of which
can lead to the damping of short wavelength
slow MHD waves, albeit at rates about two orders of
magnitude smaller than the damping rates
for the fast MHD and Alfv\'en waves at the same wavelength (Balsara 1996)}.

Estimating the corresponding length and mass scales in the dense regions of the GC 
is hampered by the significant uncertainty 
on the environmental conditions within this region, with 
$10 \le \zeta/\zeta_{CR} \le 40$, $10^7$ cm$^{-3} \le n
\le 10^8$ cm$^{-3}$, and an uncertainty as to whether or not
the relation between $n_i$ and $n$ given by Eq. (3) holds here.
In addition, the density and magnetic field strengths 
are likely to fluctuate significantly within this region
(indeed, in typical GMCs, $\delta B \sim B$).  
To keep the analysis as simple as possible,  
we plot in Figure 1 the mass of condensations 
($M_C \approx \rho\lambda_a^3$) that will
decouple from the magnetic field 
as a function of density for $\zeta/\zeta_{CR}$ = 10 (solid curve), 20
(short-dashed curve), and 40 (long dashed curve).
The dotted curve shows the Jeans mass as a function of density for $T = 25$ K.
In order to aid our analysis, we also plot the ambipolar diffusion 
time $\tau_{AD}$ as given by Eq. (4) in Figure 2.
Note that the lower branch of each curve in Fig. 1 
corresponds to the case for which Eq. (3) holds for all densities,
and the upper branch corresponds to the case where $n_i$ is
constant above $n_{crit} = 10^6$ cm$^{-3} \,(\zeta /\zeta_{CR})$.
Conversely, the upper branch of each curve in Fig. 2
corresponds to the case for which Eq. (3) holds for all densities,
and the lower branch corresponds to the case where $n_i$ is
constant above $n_{crit}$.

Our results indicate that condensations which form in
cool (25 K), dense regions of the GC molecular
clouds would exceed their Jeans limit unless the
ionization fraction is high ($\zeta \simgreat 30 \zeta_{CR}$) and
the relation between $n_i$ and $n$ given by Eq. (3) holds throughout
the core environment.  
Whether or not condensations in excess of their Jeans mass could be supported
against gravitational collapse by slow MHD waves is beyond the scope
of this paper. We note, however, that even if this were the case, the
greater densities of these condensations over the rest of the
core (due to the partial loss
of magnetic support) would then lead to lower ambipolar diffusion
times, as can be seen in Fig. 2.  
As such, support by slow MHD
waves can at best delay the gravitational collapse of these massive condensations, 
meaning that the collapse timescale would then be set by the ambipolar diffusion time
rather than the free-fall time.  

Alternatively, if condensations form with masses below their Jeans limit, 
ambipolar diffusion would first need to remove the magnetic field
globally from dense regions of the GC, a process that would take
$\sim 0.1 - 1$ Myrs.  One would then expect these regions to fragment before
collapse ensues.
It is important to note that Fig. 1 presents
an ``average" value of condensation mass expected to form in a region
with density $n$. The presence of density and magnetic 
fluctuations  expected within this highly dynamic region
would almost certainly lead to the formation of a fairly wide distribution
of condensation masses.  

This result may have important consequences for star formation 
theory. Specifically, while the molecular environment at the GC is extreme
compared to local GMC environment, the relevant physical quantities
associated with ambipolar diffusion driven star formation (mainly, the 
flux to mass ratio and ambipolar diffusion times) are similar in both
environments.  Additionally, it is intriguing that the sound speed for a $T = 25$K cloud
is $a = 0.32$ km s$^{-1}$, very near the effective sound speed associated with
typical star forming regions, as a star's initial mass may depend sensitively
on the temperature of the gas in which it is born (e.g., Adams \& Fatuzzo 1996).

\section{Conclusion}
The fact that the physical conditions at the GC are quite different
from those elsewhere in the Galaxy has often been cited as the reason
why one might expect star formation to proceed differently there than
out in the disk. But although the GC does
contain a remarkable population of massive
stars, the IMF in the Arches cluster appears to be
consistent with the universal IMF observed throughout the rest of the galaxy.

Motivated by the ongoing debate about the nature of star formation, 
we have considered here how the extreme environmental conditions
observed at the GC may affect the star formation process
within the context of the standard paradigm. 
Specifically,  we have explored the loss of magnetic support in 
dense ($n \sim 2 - 10 \times 10^7$ cm$^{-3}$), highly ionized
($\zeta \sim 1 - 4 \times 10^{-16}$ s$^{-1}$), and cool ($T \approx 25$K) molecular cores
located within the GC region. 
We find that while the higher ionization rates at the 
GC imply a strong coupling between the magnetic field and the neutral medium
for most of the GC molecular gas, the resulting effect on ambipolar diffusion is
offset by the higher densities associated with this region.  
As such, the ambipolar diffusion timescales in the dense regions of the GC
are comparable to those associated with dense
star forming cores in typical GMCs.
However, magnetic turbulence in the dense regions of the GC is suppressed
for lengthscales below $\lambda_\alpha \sim 10^{17}$ cm, 
owing to the strong
damping of Alfv\'en and fast MHD waves.
The corresponding loss of pressure is expected to lead to the formation of condensations
with masses $\sim 0.1 - 30 M_\odot$.  While this mechanism parallels the process which sets the length scale at which ambiplar diffusion
becomes significant and fragmentation or core formation is initiated in typical GMC environments
(Mouschovias 1987, 1991; Mouschovias \& Psaltis 1995), the resulting
condensations at the GC may significantly exceed their Jeans limit
($\sim 0.3 - 0.9 M_\odot$). Fragmentation of such condensations
is unlikely to yield a
core mass function similar to those observed from nearby star forming regions.  As such, 
a star formation scenario in which the IMF mimics the condensation mass function
would not be consistent with observations.  It is interesting to note,
then, that the sound speed for a 25 K molecular cloud is 0.32 km s$^{-1}$,
nearly the same as the effective sound speed associated with 
nearby star forming cores.  This result may then be taken as evidence favoring
a scenario in which stellar processes which depend primarily upon 
physical parameters such as the local sound speed set a star's initial mass
(e.g., Adams \& Fatuzzo 1996; Ciolek \& Basu 2006).    

On a final note, we note that the ambipolar diffusion time in both 
the dense regions 
of the GC and star forming cores in typical
GMCs is $\sim 1$ Myrs.  While the issue of a cloud's lifetime
remains open, recent work has established the ages of typical clouds
to be $\sim 10$ Myrs (Goldsmith et al. 2007).  This result is consistent with 
the idea that 
the effective sound crossing time yields a reasonable 
estimate for how long clouds can survive.  For a typical GMC environment, one thus
infers a cloud lifetime of $\tau_{cloud}\sim 40$ pc / (1 km/s) $\sim 40$ Myrs.
In contrast, the GC molecular clouds' lifetimes are inferred to be $\sim 1 - 3$ Myrs. 
If condensations are initially supported by slow MHD waves, their collapse
would occur on an ambipolar diffusion timescale, rather than the free-fall 
timescale.  The fact that $\tau_{AD} \sim \tau_{cloud}$ at the GC may then help
explain the low star formation efficiency in the GC region (Gordon et al. 1993).  

\vspace*{-0.3cm}
\section*{Acknowledgments}
We thank Fred Adams for useful discussions, and to the anonymous referee for 
comments that helped us significantly improve the text.
MF was partially supported by Xavier University through the
Hauck Foundation. This work was also supported by NSF
grant AST-0402502 at the University of Arizona.

\noindent Correspondence and requests for material should be addressed
to fatuzzo@xavier.edu.

\begin{figure}[p]
\begin{center}
\rotatebox{0}{\resizebox{2.8in}{!}{\includegraphics{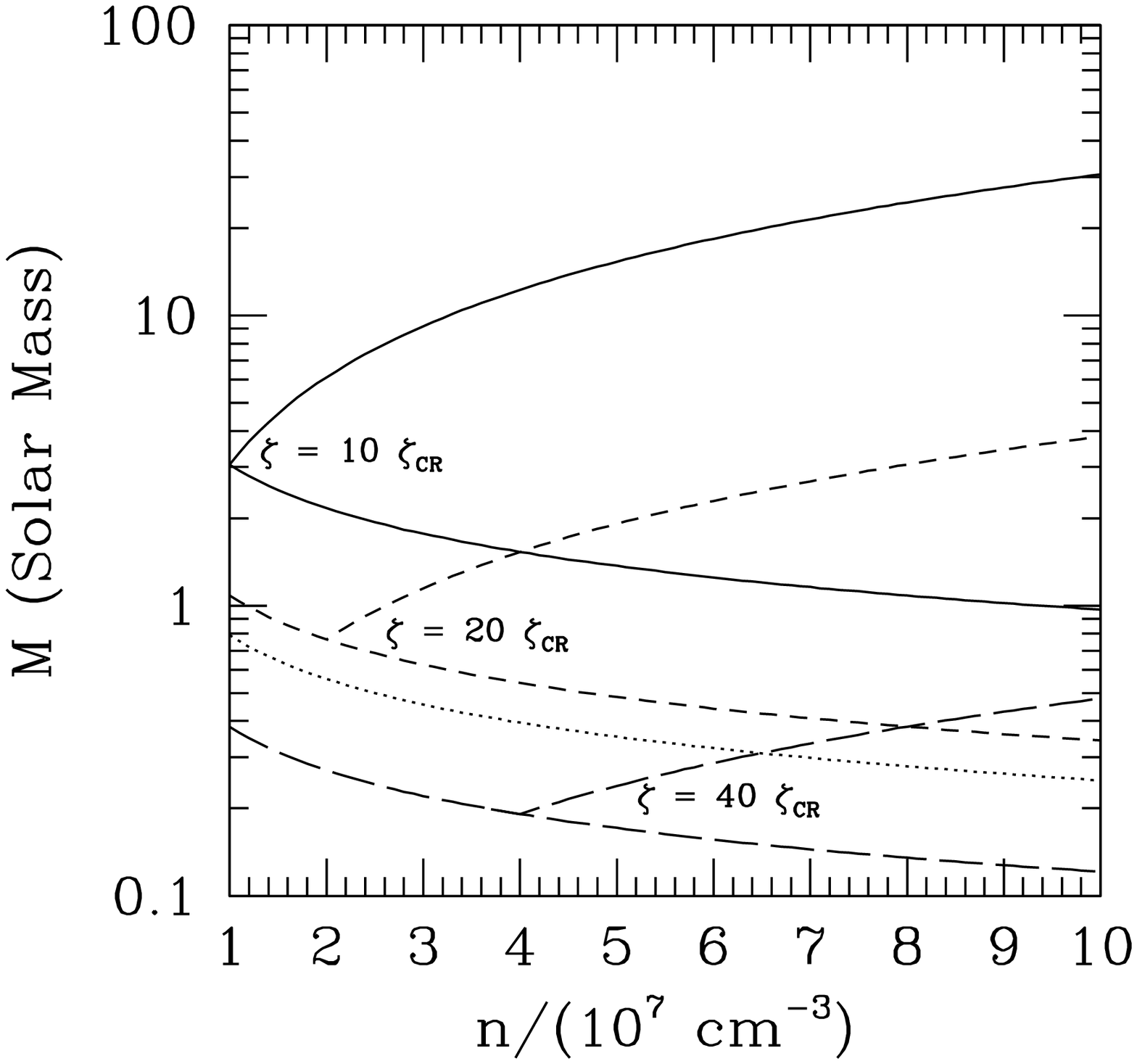}}}
\caption{\footnotesize {\bf Condensation Mass as a Function of Density.}
The mass of condensations that form as a result of magnetic decoupling due
to damping of Alfv\'en and fast MHD waves as a function of density for three
different ionization rates:  10 $\zeta_{CR}$ (solid curve); 20 $\zeta_{CR}$ 
(short dashed curve); and 40 $\zeta_{CR}$ (long dashed curve).  At densities
above $n_{crit} = (\zeta/\zeta_{CR}) \, 10^6$ cm$^{-3}$, the lower branch 
of each curve corresponds to the case where $n_i$ is set through the scaling
given by Eq. (3) for all $n$, and the upper branch corresponds to 
the case for which $n_i$ is assumed constant for densities above $n_{crit}$.   
The dotted curve indicates the Jeans mass as a function of density for
$T = 25$ K.}
\label{figure1}
\end{center}
\end{figure}

\begin{figure}[p]
\begin{center}
\rotatebox{0}{\resizebox{2.8in}{!}{\includegraphics{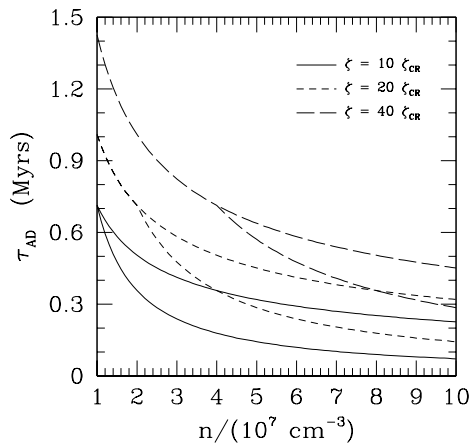}}}
\caption{\footnotesize {\bf Ambipolar Diffusion Time as a Function of Density.}
The ambipolar diffusion time as a function of density as given
by Eq. [4] for  three
different ionization rates:  10 $\zeta_{CR}$ (solid curve); 20 $\zeta_{CR}$ 
(short dashed curve); and 40 $\zeta_{CR}$ (long dashed curve).
At densities
above $n_{crit} = \zeta/\zeta_{CR} \, 10^6$ cm$^{-3}$, the upper branch 
of each curve corresponds to the case where $n_i$ is set through the scaling
given by Eq. (3) for all $n$, and the lower branch corresponds to 
the case for which $n_i$ is assumed constant for densities above $n_{crit}$.}
\label{figure1}
\end{center}
\end{figure}

\end{document}